\documentclass{pasj00}

\begin{document}
\SetRunningHead{T.Nagayama et al.}{Astrometry of ON1 with VERA}
\Received{2010/06/18}
\Accepted{2010/12/01}

\title{Astrometry of Galactic Star-Forming Region Onsala 1 with VERA:
        Estimation of Angular Velocity of Galactic Rotation at Sun}

\author{Takumi    \textsc{Nagayama},\altaffilmark{1}
         Toshihiro \textsc{Omodaka},\altaffilmark{2}
         Akiharu   \textsc{Nakagawa},\altaffilmark{2}
         Toshihiro \textsc{Handa},\altaffilmark{3} \\
         Mareki    \textsc{Honma},\altaffilmark{1}
         Hideyuki  \textsc{Kobayashi},\altaffilmark{1}
         Noriyuki  \textsc{Kawaguchi},\altaffilmark{1}
         and
         Takeshi   \textsc{Miyaji}\altaffilmark{1}
         }
\altaffiltext{1}{Mizusawa VLBI Observatory, National Astronomical Observatory of Japan, \\
                   2-21-1 Osawa, Mitaka, Tokyo 181-8588}
\email{takumi.nagayama@nao.ac.jp}
\altaffiltext{2}{Graduate School of Science and Engineering, Kagoshima University,\\
                   1-21-35 K\^orimoto, Kagoshima, Kagoshima 890-0065}
\altaffiltext{3}{Institute of Astronomy, The Universe of Tokyo, 2-21-1 Osawa, Mitaka, Tokyo 181-0015}

%


\maketitle

\begin{abstract}
We conducted the astrometry of H$_2$O masers 
in the Galactic star-forming region Onsala 1 (ON1) 
with VLBI Exploration of Radio Astrometry (VERA).
We measured a trigonometric parallax of $0.404 \pm 0.017$ mas,
corresponding to a distance of $2.47 \pm 0.11$ kpc.
ON1 is appeared to be located near the tangent point 
at the Galactic longitude of \timeform{69.54D}.
We estimate the angular velocity of the Galactic rotation at Sun, 
the ratio of the distance from Sun to the Galactic center and 
the Galactic rotation velocity at Sun, to be 
$\Omega_0 = \Theta_0 / R_0 = 28.7 \pm 1.3$ km s$^{-1}$ kpc$^{-1}$
using the measured distance and proper motion of ON1.
This value is larger than the IAU recommended value of 
220 km s$^{-1}$/8.5 kpc = 25.9 km s$^{-1}$ kpc$^{-1}$,
but consistent with other results recently obtained with the VLBI technique.
\end{abstract}


\section{Introduction}

Very Long Baseline Interferometry (VLBI) astrometry is an
important method to measure the structure of the Milky Way Galaxy (MWG).
By measuring the accurate position of the source and its time variation,
the source distance and proper motion could be determined directly.
VLBI astrometry at 10 $\mu$as accuracy
of the Galactic H$_2$O and CH$_3$OH maser sources
with VLBI Exploration of Radio Astrometry (VERA) and Very Long Baseline Array (VLBA)
determined accurate distances at kpc-scale with the errors less than 10\%
(see for example \cite{hac06}; \cite{xu06}; \cite{hon07}).

The Galactic constants, the distance from Sun to the Galactic center ($R_0$)
and the Galactic rotation velocity at Sun ($\Theta_0$) are 
major parameters to study the structure of the MWG.
The rotation curve of the MWG and all kinematic distances of the sources in the MWG
are based directly on the Galactic constants. 
Although International Astronomical Union (IAU) has recommended to give 
the values of $R_0 = 8.5$ kpc and $\Theta_0 = 220$ km s$^{-1}$ since 1985, 
numerous recent studies report the values different from them (e.g. \cite{rei09a}).

However, observational estimation of the Galactic constants is difficult.
This is because the observational estimation of Galactic constants is 
affected by several independent assumptions; 
the peculiar motion of the source, systemic non-circular motions of both the source
and the LSR due to the spiral arm and the non-axisymmetric potential of the MWG,
and relative motion of Sun to the LSR (\cite{rei09a}; \cite{mcm10}).
To minimize these effects,
we should observe many sources located at various positions in the MWG.

The tangent point that 
the vector of the source Galactic rotation is parallel to the line of sight 
is a kinematically unique position in the MWG.
In the case that the source is located at this point and on the pure circular rotation,
the proper motion of the source depends only on the Galactic rotation of Sun, $\Theta_0$.
Therefore, we can estimate $\Theta_0$ from the measured proper motion.
We can also estimate $R_0$ from 
the measured source distance,
since the tangent point, Sun, and the Galactic center make the right triangle.
\citet{sat10} measured the parallactic distance of W51 Main/South
which is located near the tangent point, and estimated $R_0 = 8.3 \pm 1.1$ kpc
using this simple geometry.
Even if the source is not located at the tangent point, but near there,
we can estimate the ratio of Galactic constants, $\Theta_0/R_0$, 
the angular velocity of the Galactic rotation at Sun, $\Omega_0$, 
as described in Section 4.
The value of this ratio is a constraint to estimate one of
the Galactic constants from the other.
Although the IAU gives recommended values of the Galactic constants,
at least one of them should be revised, if the ratio is inconsistent to the observed value.

The radial velocity of the source located at the tangent point is equal to 
the terminal velocity.
It is defined as the extreme velocity on any line of sight at 
$b \simeq \timeform{0D}$, and described as $v_{\rm term} = \Theta - \Theta_0 \sin l$,
where $\Theta$ is the Galactic rotation velocity of the source.
Therefore, we selected the sources whose radial velocities are close 
to the terminal velocity.
One of these sources, Onsala 1 (ON1) is a massive star-forming region
located at the Galactic coordinate of $(l, b) = (\timeform{69.54D}, \timeform{-0.98D})$.
The radial velocity of ON1 is close to the terminal velocity.
The radial velocity observed in molecular lines 
is $12\pm1$ km s$^{-1}$ (\cite{bro96}; \cite{pan01}).
The terminal velocity at $l = \timeform{69.54D}$ is derived to be
14 km s$^{-1}$ from the IAU recommended value of $\Theta_0 = 220$ km s$^{-1}$ 
and assuming the flat rotation ($\Theta = \Theta_0$),
and is $15\pm5$ km s$^{-1}$ in the $l$-$v$ diagram of \citet{dam01}.
The distance of ON1 is measured to be $2.57^{+0.34}_{-0.27}$ kpc
by the 6.7 GHz CH$_3$OH maser astrometry with the EVN (\cite{ryg09}).
VLBI maps of the H$_2$O masers have been reported by \citet{nag08}. 
They found that ON1 has two clusters of H$_2$O masers (WMC1 and WMC2) 
separated by \timeform{1.6"}.

In the present study, 
we report on our successful determination of the parallax of ON1 with VERA.
This is a first step to estimate the Galactic constants and 
the angular velocity of the Galactic rotation at Sun using VERA.


\section{Observations and Reductions}

We observed H$_2$O masers in the star-forming region 
ON1 with VERA at 11 epochs spanned about two years.
The epochs are day of year (DOY) 245, 256, 290, 309, 360 in 2006,
046, 091, 129, 223 in 2007, 020, and 202 in 2008.
At each epoch, the H$_2$O $6_{16}$-$5_{23}$ maser at 
a rest frequency of 22.235080 GHz in ON1 and 
a position reference source J2010+3322 were simultaneously observed 
in a dual-beam mode for about 10 hours.
The typical on-source integration time was 6 hours
for both ON1 and J2010+3322.
J2010+3322 is listed in VLBA Calibrator Survey 2 (VCS2: \cite{fom03}).
J2010+3322 was detected with a peak flux density of 100--130 mJy
in each epochs.
The separation angle between ON1 and J2010+3322 is \timeform{1.85D}.
The instrumental phase difference between the two beams
was measured continuously during the observations by
injecting artificial noise sources into both beams (\cite{hon08a}).
Left-hand circular polarization signals were sampled with 2-bit quantization,
and filtered with the VERA digital filter unit (\cite{igu05}).
The data were recorded onto magnetic tapes at a rate of 1024 Mbps,
providing a total bandwidth of 256 MHz, which consists of $16\times16$ MHz IF channels.
One IF channel was assigned to ON1, and the other 15 IF
channels were assigned to J2010+3322, respectively.
Correlation processing was carried out on the Mitaka FX correlator.
The frequency and velocity resolutions for ON1
were 31.25 kHz and 0.42 km s$^{-1}$, respectively.

Data reduction was conducted using 
the NRAO Astronomical Image Processing System (AIPS).
An amplitude calibration was performed using 
the system noise temperatures during the observations.
For phase-referencing, a fringe fitting was
made using the AIPS task FRING on J2010+3322 
with a typical integration time of 2 min.
The solutions of the fringe phases, group delays, 
and delay rates were obtained every 30 sec.
These solutions were applied to 
the data of ON1 in order to calibrate the visibility data.
Phase and amplitude solutions obtained from self-calibration of J2010+3322
were also applied to ON1.
Visibility phase errors caused by the Earth's atmosphere were calibrated
based on GPS measurements of the atmospheric zenith delay which occurs
due to tropospheric water vapor (\cite{hon08b}).
After the calibration, 
we made spectral-line image cubes using the AIPS task IMAGR around masers with
$1024\times1024$ pixels of size 0.05 milliarcsecond (mas).
The typical size of the synthesized beam was $1.3\times0.9$ mas 
with position angle of \timeform{-40D}.
The rms noises for each images were approximately 0.1--1 Jy beam$^{-1}$.
The signal-to-noise ratio of 7 was adopted as the detection criterion.

In the single-beam VLBI imaging of ON1
(without phase-referencing to the J2010+3322 in the other beam),
group delays solved on J2025+3343 in the same beam were applied.
The fringe fitting was done using 
one of the brightest emissions at LSR velocity of 14.9 km s$^{-1}$ in ON1,
and the solutions are applied to all velocity channels.
The data reductions for single-beam imaging are the same method with 
previous JVN observations of ON1 (\cite{nag08}). 


\section{Results}

\subsection{Overall Properties of H$_2$O Masers in ON1}

\begin{figure}
  \begin{center}
    \FigureFile(80mm,80mm){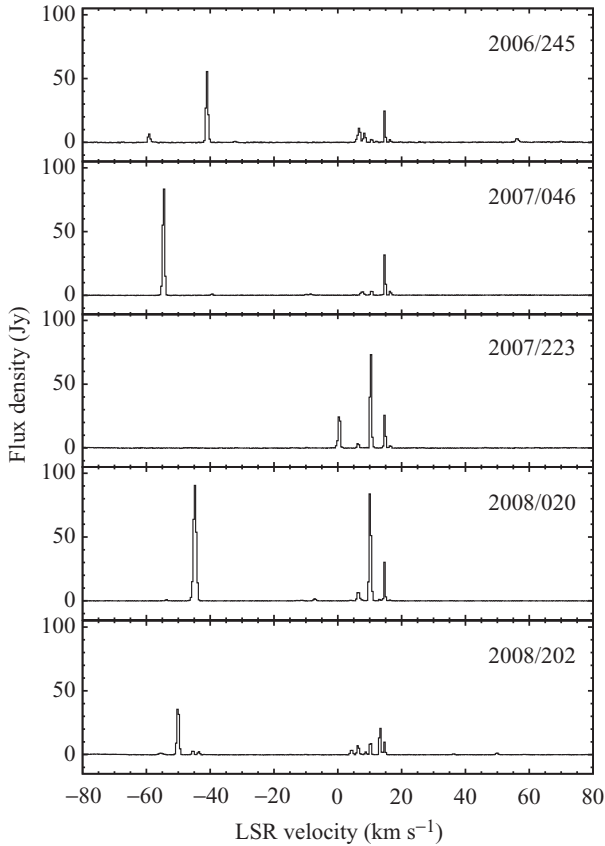}
  \end{center}
  \caption{Scalar-averaged cross-power spectra of 
the ON1 H$_2$O masers obtained with Mizusawa-Iriki baseline.}
  \label{fig:1}
\end{figure}

Figure \ref{fig:1} shows the scalar-averaged cross-power spectra
of ON1 H$_2$O masers in approximately half a year intervals.
They are obtained with the Mizusawa-Iriki baseline,
and averaged for the total observational time.
There are low-velocity components ($v_{\rm LSR} = -7$ to 17 km s$^{-1}$)
near the systemic velocity of ON1 ($v_{\rm LSR} = 12 \pm 1$ km s$^{-1}$;
\cite{bro96}),
and high-velocity blue-shifted ($v_{\rm LSR} = -57$ to $-32$ km s$^{-1}$)
and red-shifted ($v_{\rm LSR} = 55$ to 70 km s$^{-1}$) components.
The most of the low-velocity components are detected over one year.
The high-velocity components are time-variable and 
not persistent for more than half a year.

\begin{figure*}
  \begin{center}
    \FigureFile(120mm,120mm){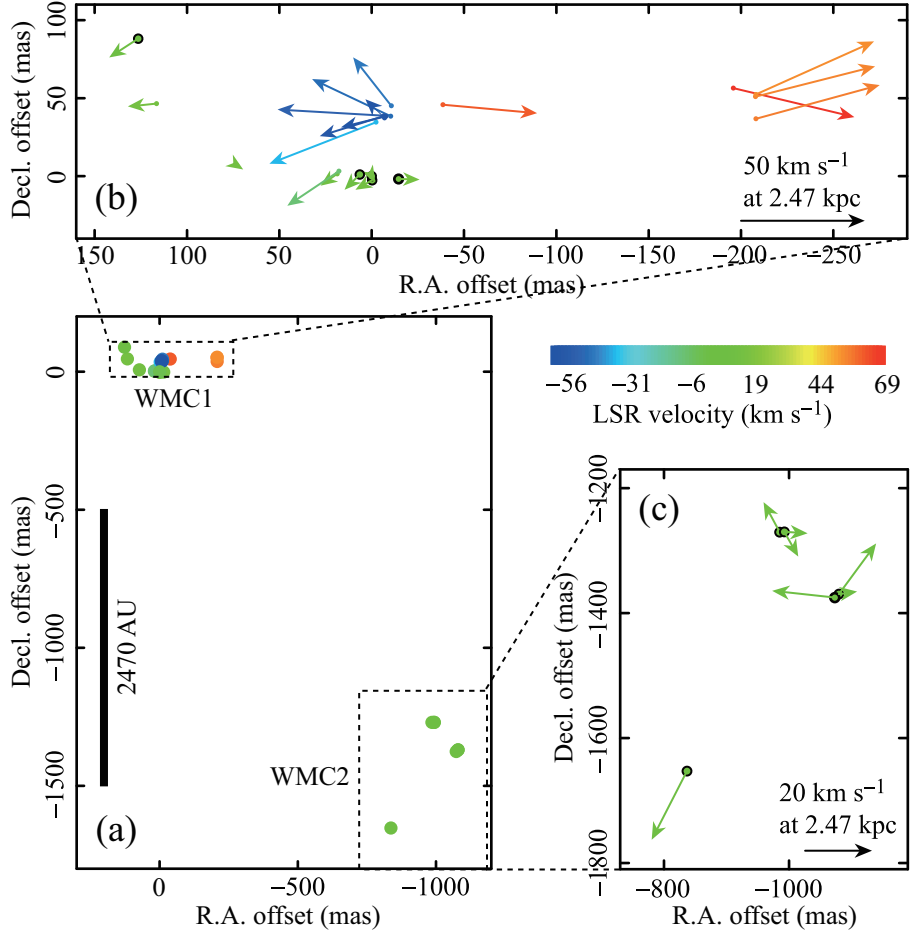}
  \end{center}
  \caption{(a): Distributions of H$_2$O masers in ON1.
            The color index denotes the LSR velocity range from $-56.8$ to 69.6 km s$^{-1}$,
            where 28 features are located.
            The map origin is located at the position of the reference maser feature at 
            $v_{\rm LSR} = 14.9$ km s$^{-1}$, and
            $(\alpha, \delta)_{\rm J2000} = (\timeform{20h10m09.20454s}, \timeform{31D31'36.1012"})$
            in 2006/245.
            (b), (c): Close-up to the two maser clusters (WMC1 and WMC2) with 
            internal proper-motion vectors.
            The maser features measured the parallax and proper motions 
            are highlighted by black circle.
            }
  \label{fig:2}
\end{figure*}

We detected 28 H$_2$O maser features for more than two epochs.
Figure \ref{fig:2} shows the distributions and internal motions of these features.
The masers are distributed with approximately 
$\timeform{1"} \times \timeform{2"}$ area,
which is consistent with the previous VLBI observations (\cite{nag08}).
The difference between the present and previous observations
is the difference in time intervals of the epochs.
By observing with the shorter time intervals of 1--2 months,
the number of the maser features which we can trace the internal motions 
in the present observations increased to the twice of the previous ones.
ON1 has two clusters of H$_2$O maser features located at 
$(x, y) \simeq (\timeform{0.0"}, \timeform{0.0"})$
and (\timeform{-0.9"}, \timeform{-1.4"}).
They are named as WMC1 and WMC2 by \citet{nag08}, respectively.

The internal motions of WMC1 exhibit a bipolar outflow structure
in the east-west direction.
The blue- and red-shifted features show the high expansion velocity of
$\simeq$ 70 km s$^{-1}$ in three-dimensional velocity.
The low velocity features with $v_{\rm LSR} = -7$ to 17 km s$^{-1}$
show low expansion velocity of $\simeq$ 10 km s$^{-1}$.
In the previous observations,
the blue- and red-shifted features were distributed 
within two small regions of approximately $15\times 15$ mas 
separated by $\simeq 200$ mas, 
and were not detected outside of these regions (\cite{nag08}).
They are two clusters of the blue-shifted features  
around $(x, y) \simeq (-5, 45)$ mas and the red-shifted features 
around $(-200, 50)$ mas
in Figure \ref{fig:2}(b).
\citet{nag08} suggest that a driving source of H$_2$O masers is 
located near the midpoint of these two clusters.
However, we detected a new red-shifted feature at 
$(x, y) \simeq (-40, 45)$ mas.
Therefore, the driving source would be located between
the cluster of blue-shifted features and a new red-shifted feature.
In WMC2, only the low velocity features ($v_{\rm LSR} = 6$ to 14 km s$^{-1}$)
are detected.
Their internal motions did not show a systematic,
and thus masers in WMC2 are not likely to originate from WMC1.
We confirmed that there are two driving sources of H$_2$O masers in ON1.

\subsection{Parallax and Proper Motion}

\begin{figure*}
  \begin{center}
    \FigureFile(160mm,160mm){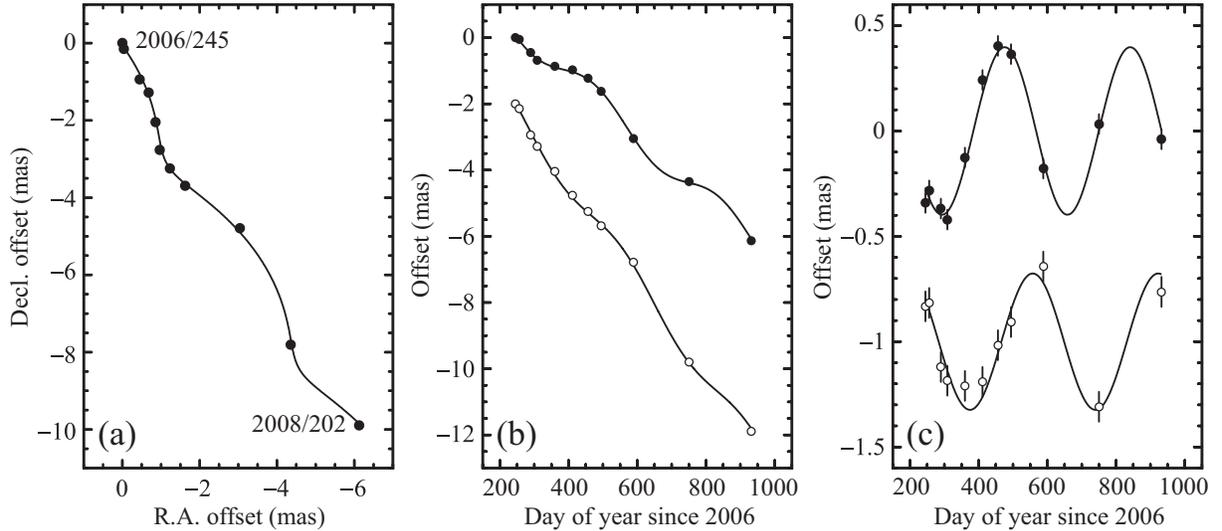}
  \end{center}
  \caption{Parallax and proper motion data and fits for 
  the maser feature of $v_{\rm LSR} = 14.9$ km s$^{-1}$.
  (a): Positions on the sky with first and last epochs labeled.
  Solid line indicates the parallax and proper motion fit.
  (b): $x$ (filled circles) and $y$ (open circles) position offsets versus time.
  Solid line indicates the parallax and proper motion fit.
  The $y$ data have been offset from the $x$ data for clarity.
  (c): same as (b) panel, expect the proper motion fit has been removed,
  allowing the effects of only the parallax to be seen.}
  \label{fig:3}
\end{figure*}

Absolute motions of H$_2$O masers 
to the extragalactic position reference source 
are expressed by the sum of a linear motion and the trigonometric parallax.
We conducted monitoring observations of H$_2$O masers for about two years,
and their absolute motions were successfully obtained by referencing to
the position reference source, J2010+3322.
Figure \ref{fig:3} shows positional variations of one of the brightest
maser feature at $v_{\rm LSR} = 14.9$ km s$^{-1}$.
The positional variations show 
systematic sinusoidal modulation with a period of one year caused by the parallax.

In order to obtain the parallax and proper motion,
we use the positions of ten maser features detected over half a year.
We conducted a combined parallax fit, 
in which the positions of ten features are fitted simultaneously
with one common parallax but different proper motions and position offsets for each spot.
For this fit, we used errors of 0.057 and 0.082 mas
in right ascension and declination, respectively, in quadrature
to the formal fitting errors.
This resulted in $\chi^2$ per degree of freedom values of unity
for both the right ascension and declination data.
The resulting parallax is $0.404\pm0.012$ mas.
Table \ref{tab:1} shows the results of the combined parallax fit.

To confirm the consistency of the parallax motion for each maser feature,
we also estimate the parallax individually.
In Table \ref{tab:1}, 
we also show the obtained parallaxes using individual fitting for each maser feature.
The obtained parallaxes of ten maser features are
consistent with each other, from 0.382 to 0.428 mas,
and give the similar result of $0.404\pm0.012$ mas.

Combining the results of ten maser features can lead
to underestimation of the parallax uncertainty,
in the case that the measurements are correlated among different maser features.
Random-like errors such as map noise and maser feature structure variation
would not be correlated among different maser features.
However, systemic errors (e.g. from the correlator model or the atmosphere) 
would affect all maser features in one epoch in a very similar way.
The conservative approach would be that the uncertainty is not reduced
when the combined fitting of ten maser features are made.
In this approach, the uncertainty would be estimated to be 0.021 mas 
from the smallest uncertainty of ID 3 maser feature.
However, the results of ten maser features are not entirely correlated,
since the obtained parallaxes from individual fitting are not completely the same value.
Therefore, we estimate the final uncertainty of the parallax
by taking the middle between the uncertainties of 0.012 and 0.021 mas,
and obtain $0.404\pm0.017$ mas, which we adopt for the parallax of ON1.

\begin{figure}
  \begin{center}
    \FigureFile(80mm,80mm){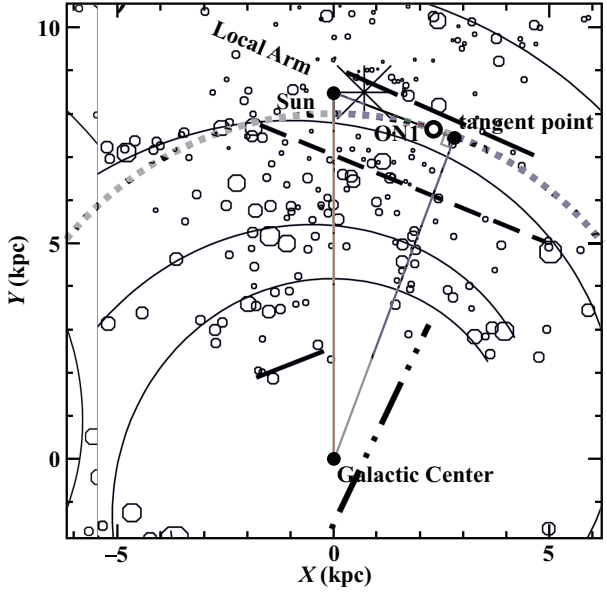}
  \end{center}
  \caption{Position of the ON1 in the MWG.
            The background is the four spiral arm structure of the MWG (\cite{rus03}).
            The thick lines sketched by \citet{rus03} show 
            the local arm feature (long dashed line),
            the bar orientation and length (dashed-dot-dot line) from \citet{eng99},
            the expected departure from a logarithmic spiral arm observed
            for the Sagittarius-Carina arm (short dashed line),
            and a feature certainly linked to the three-kpc arm (solid line).
            }
  \label{fig:4}
\end{figure}

The obtained parallax corresponds 
to a source distance of $2.47 \pm 0.11$ kpc.
This distance is consistent with 
the 6.7 GHz CH$_3$OH maser parallax corresponding to 
$2.57^{+0.34}_{-0.27}$ kpc measured by \citet{ryg09},
but it is obtained with 2 times higher accuracy.
Figure \ref{fig:4} shows the position of ON1 in the MWG
which is determined from the distance of $2.47 \pm 0.11$ kpc,
the longitude of $l =$ \timeform{69.54D} and $R_0 = 8.5$ kpc.
ON1 is appeared to be located on 
the Local arm and near the tangent point at $l = \timeform{69.54D}$.

The absolute proper motion of a maser feature is 
the sum of the internal motion of the maser feature, 
the Galactic rotation, the Solar motion, and the peculiar motion of the source.
All motions expect the internal motion are common to all maser features.
Therefore, the average of the absolute proper motions
of maser features should give the systemic proper motion of the whole source,
if the internal motions are randomized well.
We consider that it is valid for ON1
because the averaged radial velocity of ten maser features is 11.1 km s$^{-1}$,
and it is consistent with the systemic radial velocity derived from the
associated molecular cloud ($v_{\rm LSR} = 12 \pm 1$ km s$^{-1}$; \cite{bro96}).
The absolute proper motions of ten maser features are range
from $-3.89$ to $-2.05$ mas yr$^{-1}$ in right ascension and
from $-5.58$ to $-3.15$ mas yr$^{-1}$ in declination.
From the average (the arithmetic mean) of these values 
the systemic proper motion of ON1 is derived to be
$(\mu_{\alpha} \cos \delta, \mu_{\delta}) = 
(-3.10 \pm 0.18, -4.70 \pm 0.24)$ mas yr$^{-1}$,
where the error is the standard error of the mean.
Although there is a difference of approximately 1 mas yr$^{-1}$ 
between the averaged proper motions of WMC1 and WMC2, 
it would be the relative stellar motion 
since WMC1 and WMC2 are associated with the different sources.
Averaging the motions of multiple sources can also help to 
determine the systemic motion of ON1.
We compared to the proper motion obtained by \citet{ryg09}.
\citet{ryg09} measured the proper motions of ON1 
using the two background sources, J2003+3034 and J2009+3049,
and obtained the different proper motions between them 
because of their apparent movements.
Our derived systemic proper motion is close to the proper motion
obtained using J2009+3049.

We convert the proper motion of 
$(\mu_{\alpha} \cos \delta, \mu_{\delta}) = 
(-3.10 \pm 0.18, -4.70 \pm 0.24)$ mas yr$^{-1}$
to one with respect to LSR
using the Solar motion in the traditional definition of 
$(U_{\odot}, V_{\odot}, W_{\odot}) = (10.3, 15.3, 7.7)$ km s$^{-1}$.
In the previous studies for the astrometry of the Galactic maser sources,
the Solar motion of $(U_{\odot}, V_{\odot}, W_{\odot}) = (10.00, 5.25, 7.17)$ km s$^{-1}$
based on the HIPPARCOS satellite data (\cite{deh98}) is widely used.
However, \citet{sch10} suggest that the value of $V_{\odot}$ 
determined by \citet{deh98} is underestimated by $\sim 7$ km s$^{-1}$,
and show the value which is close to the value of the traditional definition
used in this study.
Using the Galactic coordinate of ON1 $(l, b) = (\timeform{69.54D}, \timeform{-0.98D})$,
the proper motion with respect to LSR projected to the direction of $l$ and $b$ 
is calculated to be $(\mu_l, \mu_b) = (-6.00 \pm 0.22, 0.69 \pm 0.20)$ mas yr$^{-1}$.
This proper motion corresponds to a velocity of
$(v_l, v_b) = (-70.2 \pm 2.6, 8.1 \pm 2.3)$ km s$^{-1}$.

\begin{table*}
  \caption{The best-fit values of parallax and proper motions 
            for ten H$_2$O maser features in ON1.}
  \label{tab:1}
  \begin{center}
    \begin{tabular}{rrrrrrrr}
      \hline
                                                      &
      \multicolumn{1}{c}{$\Delta \alpha$}          &
      \multicolumn{1}{c}{$\Delta \delta$}          &
      \multicolumn{1}{c}{$v_{\rm LSR}$}              &
      \multicolumn{1}{c}{Detected}                   &
      \multicolumn{1}{c}{$\pi$}                      &
      \multicolumn{1}{c}{$\mu_\alpha \cos \delta$} &
      \multicolumn{1}{c}{$\mu_\delta$}              \\
      \multicolumn{1}{c}{ID}              &
      \multicolumn{1}{c}{(mas)}           &
      \multicolumn{1}{c}{(mas)}           &
      \multicolumn{1}{c}{(km s$^{-1}$)}   &
      \multicolumn{1}{c}{epochs}          &
      \multicolumn{1}{c}{(mas)}           &
      \multicolumn{1}{c}{(mas yr$^{-1}$)} &
      \multicolumn{1}{c}{(mas yr$^{-1}$)} \\
      \hline
 1 &    126.5  &     88.1  &  7.7 & 11111111100 & $0.389 \pm 0.036$ & $-2.65 \pm 0.09$ & $-5.58 \pm 0.12$ \\
 2 &      6.5  &      1.1  & 10.2 & 00011111111 & $0.428 \pm 0.025$ & $-2.73 \pm 0.05$ & $-5.44 \pm 0.08$ \\
 3 &      0.0  &      0.0  & 14.9 & 11111111111 & $0.407 \pm 0.021$ & $-3.42 \pm 0.03$ & $-5.30 \pm 0.04$ \\
 4 &    $-0.2$ &    $-2.6$ & 14.4 & 01111111110 & $0.420 \pm 0.041$ & $-3.32 \pm 0.05$ & $-5.21 \pm 0.07$ \\
 5 &   $-14.6$ &    $-1.8$ & 16.5 & 11111111111 & $0.382 \pm 0.031$ & $-3.89 \pm 0.07$ & $-4.94 \pm 0.09$ \\
 6 &  $-836.8$ & $-1653.0$ & 10.6 & 11111111111 & $0.421 \pm 0.066$ & $-3.20 \pm 0.09$ & $-5.02 \pm 0.12$ \\
 7 &  $-985.1$ & $-1270.3$ & 14.0 & 11111111000 & $0.390 \pm 0.055$ & $-3.51 \pm 0.10$ & $-4.08 \pm 0.14$ \\
 8 &  $-994.3$ & $-1269.9$ &  6.4 & 11111100000 & $0.382 \pm 0.057$ & $-3.71 \pm 0.15$ & $-3.15 \pm 0.21$ \\
 9 & $-1073.2$ & $-1375.1$ &  8.5 & 11111110000 & $0.399 \pm 0.039$ & $-2.05 \pm 0.11$ & $-3.72 \pm 0.16$ \\
10 & $-1079.9$ & $-1369.5$ &  7.3 & 00111111110 & $0.416 \pm 0.065$ & $-2.51 \pm 0.15$ & $-4.58 \pm 0.20$ \\
\hline
      \multicolumn{5}{c}{Combined fit}       &
      $0.404\pm0.012$           &
                                &
                                \\
\hline
      \multicolumn{5}{c}{Average} &
                                &
      $-3.10 \pm 0.18$          &
      $-4.70 \pm 0.24$          \\
\hline
\multicolumn{8}{@{}l@{}} {\hbox to 0pt{\parbox{150mm}{\footnotesize
Column (2), (3): Offsets relative to the positon of the maser feature at 
       $v_{\rm LSR} = 14.9$ km s$^{-1}$, and 
       $(\alpha, \delta)_{\rm J2000} = (\timeform{20h10m09.20454s}, \timeform{31D31'36.1012"})$
       in 2006/245.\\
Column (4): LSR velocity. \\
Column (5): Detected epochs: `1' for detection, and `0' for non-detection.\\
Column (6): Parallax estimates.\\
Column (7), (8): Motions on the sky along the right ascension and declination.
}\hss}}
    \end{tabular}
  \end{center}
\end{table*}

\section{Discussion}

As shown in section 1,
the source at the tangent point has a kinematically unique property
and one of the key objects to study the kinematics and geometry of the MWG.
Due to a symmetric geometry,
the radial velocity of a source at the 
tangent point with a pure circular rotation is equal to the terminal velocity.

\begin{figure*}
  \begin{center}
    \FigureFile(160mm,80mm){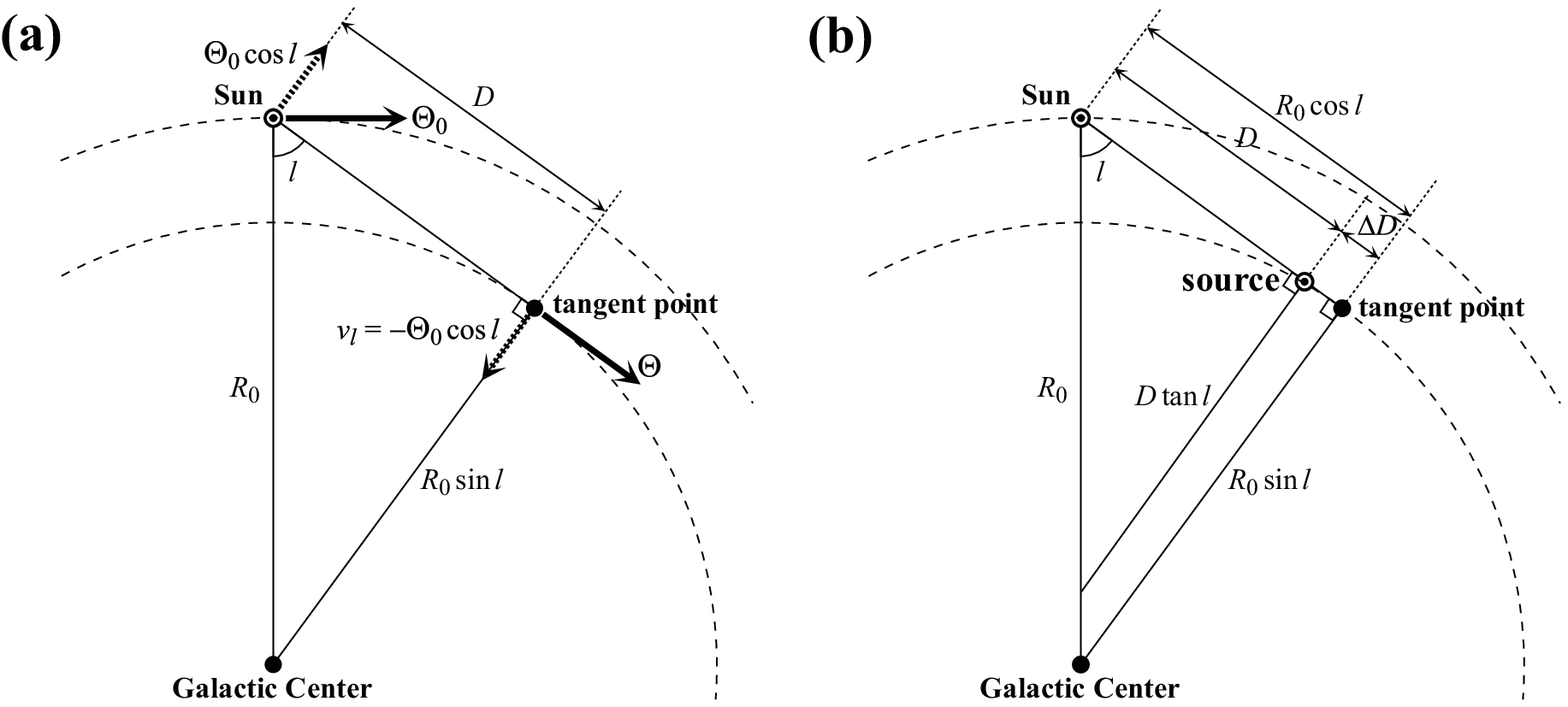}
  \end{center}
  \caption{The geometry of the Galactic center, Sun, the tangent point, and the source.
           (a): The geometry in the case that the source is located at the tangent point.
           (b): The geometry in the case that there is a offset between the source and the tangent point.}
  \label{fig:5}
\end{figure*}

The radial velocity of ON1 observed in molecular lines
is $12\pm1$ km s$^{-1}$ (\cite{bro96}; \cite{pan01}),
The terminal velocity at $l = \timeform{69.54D}$ is $15 \pm 5$ km s$^{-1}$ (\cite{dam01}).
They are the same within the error.
This suggests that ON1 is located at the tangent point.
In the case that the source is located exactly at the tangent point, 
and it is on pure circular rotation,
the source, Sun, and the Galactic center make the right triangle,
and the source proper motion on the sky depends only on the Galactic rotation of Sun.
This geometry is shown in Figure \ref{fig:5}(a).
Therefore, $R_0$ and $\Theta_0$ are determined 
from the observed distance to the source, $D$, and
the proper motion on the sky along the Galactic plane, $v_l$, as
\begin{eqnarray}
R_0       & = & D / \cos l 
\label{equ:1} \\
\Theta_0 & = & - v_l / \cos l.
\label{equ:2}
\end{eqnarray}
The Galactic constants are estimated to be $R_0 = 7.1\pm0.3$ kpc 
and $\Theta_0 = 201\pm7$ km s$^{-1}$, respectively,
from $D = 2.47\pm0.11$ kpc and $v_l = -70.2\pm2.6$ km s$^{-1}$.
These values are approximately 10--20\% smaller than 
the IAU recommended values of $R_0 = 8.5$ kpc and $\Theta_0 = 220$ km s$^{-1}$,
the recently estimated values of 
$R_0 = 8.4 \pm 0.6$ kpc and $\Theta_0 = 254 \pm 16$ km s$^{-1}$ (\cite{rei09a}),
and $R_0 = 8.3 \pm 1.1$ kpc estimated using
the parallax measurement of W51 Main/South located near the tangent point (\cite{sat10}).

However, our estimated $R_0$ and $\Theta_0$ would 
not be inconsistent with the previous estimates.
This is because our estimation is affected 
by the ambiguities of the two assumptions  
that ON1 is on the pure circular rotation, 
and located exactly at the tangent point.
We evaluate these effects.
Since the velocity deviation of the molecular cloud in ON1
is estimated to be 3--8 km s$^{-1}$ from the 
full width at half maximum of the profiles in molecular lines
(\cite{zhe85}; \cite{has90}; \cite{bro96}),
the maser source in ON1 may have the peculiar motion of 
the similar velocity.
If we allow for this peculiar motion of 3--8 km s$^{-1}$,
the errors of our estimated $R_0$ and $\Theta_0$ are larger than 10--20\%.
In the case that we use the IAU recommended values of $R_0 = 8.5$ kpc 
and $\Theta_0 = 220$ km s$^{-1}$, and assume the flat rotation, 
the peculiar motion is estimated to be 
$(U', V', W') = (7.0\pm2.5, -1.8\pm1.3, 8.1\pm2.3)$ km s$^{-1}$
from the measured distance, proper motion, and radial velocity 
using the method shown in \citet{rei09a}.
Here, $U'$ is the velocity component toward the Galactic Center,
$V'$ is the component in the direction of the Galactic rotation,
$W'$ is the component toward the north Galactic pole.
In the case that we use different Galactic constants of 
$R_0 = 8.4$ kpc and $\Theta_0 = 254$ km s$^{-1}$ (\cite{rei09a}), 
the peculiar motion is estimated to be
$(U', V', W') = (-3.7\pm2.5, -3.2\pm1.3, 8.1\pm2.3)$ km s$^{-1}$.
In both cases, the peculiar motion velocity is in the range of
the velocity deviation of the molecular cloud.
There may be a velocity difference of approximately 5 km s$^{-1}$ between 
the radial velocity and the terminal velocity, considering their errors.
If this velocity difference is due to the offset from the tangent point to ON1,
this offset is estimated to be approximately $\pm 1.7$ kpc.
If ON1 is not located exactly at the tangent point and 
there is a offset of $\pm 1.7$ kpc between them,
the errors of our estimated $R_0$ and $\Theta_0$ increase to approximately 70\%. 

\begin{figure*}
  \begin{center}
    \FigureFile(160mm,80mm){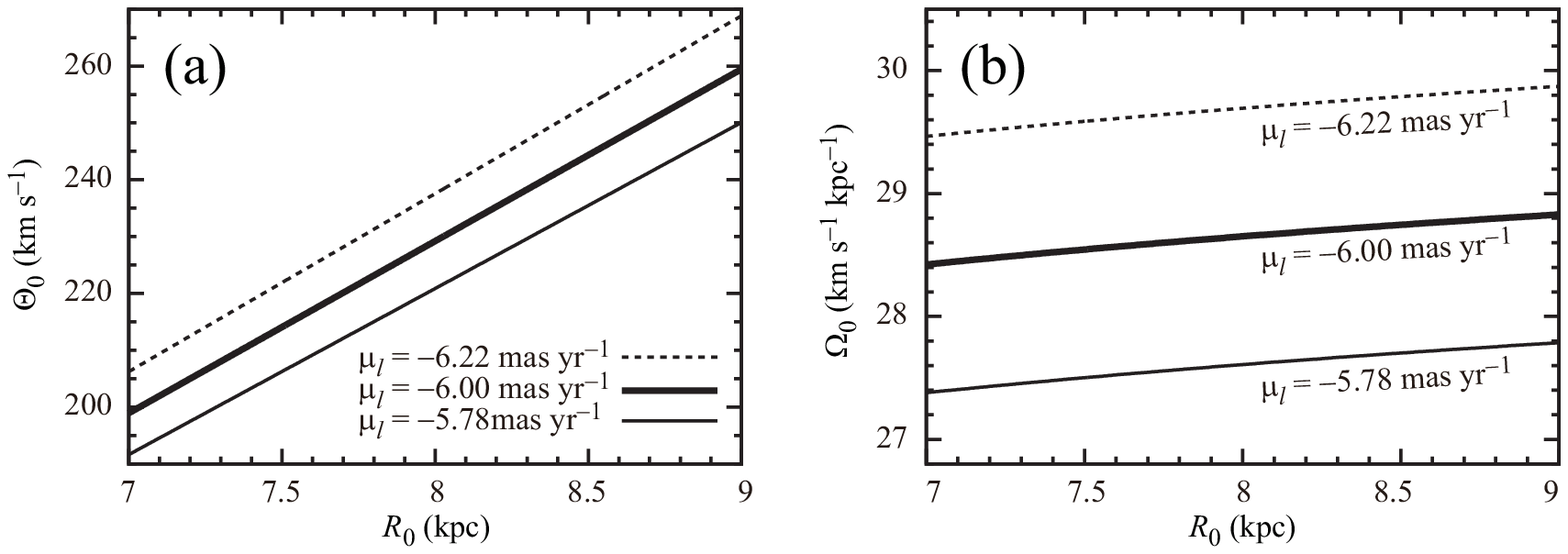}
  \end{center}
  \caption{(a): The relation of $R_0$ and $\Theta_0$ shown in the equation (\ref{equ:5}).
           (b): The relation of $R_0$ and $\Theta_0 / R_0$ shown in the equation (\ref{equ:6}).}
  \label{fig:6}
\end{figure*}

Although the estimated values of the Galactic constants are 
strongly affected by the assumption of the source location in the MWG.
However we found that the ratio of Galactic constants, $\Theta_0/R_0$, 
the angular velocity of Galactic rotation at Sun, $\Omega_0$, 
can be estimated with small ambiguity.
In the case that the source is on pure circular rotation 
at any position in the Galactic disk,
the radial and tangential velocities of the source 
can be written as
\begin{eqnarray}
v_r &=& \left( \frac{\Theta}{R} - \frac{\Theta_0}{R_0} \right) R_0 \sin l,
\label{equ:3} \\
v_l &=& \left( \frac{\Theta}{R} - \frac{\Theta_0}{R_0} \right) R_0 \cos l - \frac{\Theta}{R}D.
\label{equ:4}
\end{eqnarray}
From these equations, the relation between the $\Theta_0$ and $R_0$ is obtained to be
\begin{eqnarray}
\Theta_0 & = & \left[ -\frac{v_l}{D} + v_r \left( \frac{1}{D \tan l} - \frac{1}{R_0 \sin l} \right) \right] R_0 
\nonumber \\
          & = & \left[ -a_0 \mu_l + v_r \left( \frac{1}{D \tan l} - \frac{1}{R_0 \sin l} \right) \right] R_0,
\label{equ:5} 
\end{eqnarray}
where $a_0$ is a conversion constant from a proper motion to a linear velocity
(4.74 km s$^{-1}$ mas$^{-1}$ yr kpc$^{-1}$).
The equation (\ref{equ:5}) is graphed in Figure \ref{fig:6}(a)
using the observed values of $D = 2.47 \pm 0.11$ kpc, 
$\mu_l = -6.00 \pm 0.22$ mas yr$^{-1}$, 
and $v_{r} = 12 \pm 1$ km s$^{-1}$.
We found that the slope in Figure \ref{fig:6}(a) 
is a nearby constant at $7 \leq R_0 \leq 9$ kpc.
The slope yields the ratio $\Theta_0/R_0$, which is described as
\begin{eqnarray}
\frac{\Theta_0}{R_0} 
          & = & -\frac{v_l}{D} + v_r \left( \frac{1}{D \tan l} - \frac{1}{R_0 \sin l} \right)
\nonumber \\ 
          & = & -a_0 \mu_l + v_r \left( \frac{1}{D \tan l} - \frac{1}{R_0 \sin l} \right),
\label{equ:6} 
\end{eqnarray}
The equation (\ref{equ:6}) is graphed in Figure \ref{fig:6}(b).
The ratio is estimated to be $\Theta_0 / R_0 = 28.7 \pm 1.3$ km s$^{-1}$ kpc$^{-1}$
using the above observed values, and $7 \leq R_0 \leq 9$ kpc.
The error of $\Theta_0 / R_0$ mainly depends on that of $\mu_l$.
The errors of $\Theta_0 / R_0$ depend on those of 
$v_r$ and $D$ are $\pm 0.02$ and $\pm 0.05$ km s$^{-1}$ kpc$^{-1}$, respectively, 
and they can be neglected in this estimation.
This is because that $D \tan l \simeq R_0 \sin l$ 
in the case that the source is located near the tangent point (see Figure \ref{fig:5}(b)).
We adopted $7 \leq R_0 \leq 9$ kpc in this estimation.
This means that ON1 is located within $-0.02$--$0.68$ kpc from the tangent point,
because the offset from the tangent point is written as $\Delta D = R_0 \cos l -D$. 

Our estimated ratio $\Theta_0 / R_0 = 28.7 \pm 1.3$ km s$^{-1}$ kpc$^{-1}$
is close to the value of $27.3 \pm 0.8$ km s$^{-1}$ kpc$^{-1}$
obtained from the parallax and proper motion measurement of ON2N 
which is located on the Solar circle (\cite{and10}, submitted to PASJ).
Our estimated ratio is consistent with the 
value of $\Theta_0 / R_0 = 29.45 \pm 0.15$ km s$^{-1}$
estimated from the proper motion measurement of Sgr A* (\cite{rei04}).
\citet{rei04} adopt the Solar motion 
in the direction of Galactic rotation of $V_{\odot} = 5.25$ km s$^{-1}$ (\cite{deh98}).
We adopt the Solar motion in the traditional definition of 
$(U_{\odot}, V_{\odot}, W_{\odot}) = (10.3, 15.3, 7.7)$ km s$^{-1}$ (see subsection 3.2).
The error of our estimated ratio depends on that of $V_{\odot}$
is estimated to be $\pm 1.4$ km s$^{-1}$ kpc$^{-1}$ from $V_{\odot} = 15 \pm 10$ km s$^{-1}$.
Even if we consider this,
our estimated value is larger than the IAU recommended value of 
220 km s$^{-1}$/8.5 kpc = 25.9 km s$^{-1}$ kpc$^{-1}$.

We can find the other sources which is expected to be located at the tangent point 
from the Arcetri catalog of H$_2$O maser (\cite{val01}).
Therefore, we will observe these sources with VERA,
in order to determine the Galactic constants and 
the angular rotation velocity of the Galactic rotation at Sun.

\bigskip

We are grateful to an anonymous referee for valuable comments and suggestions.
We thank to the staff members of all the VERA stations
for their assistances in the observations.




\begin{thebibliography}{}
\bibitem[Ando et al.(2010)]{and10}
Ando, K., et al. 2010, submitted to PASJ 
\bibitem[Bronfman et al.(1996)]{bro96} 
Bronfman, L., Nyman, L.-A., \& May, J.\ 1996, \aaps, 115, 81 
\bibitem[Dame et al.(2001)]{dam01} Dame, T.~M., Hartmann, D., 
\& Thaddeus, P.\ 2001, \apj, 547, 792 
\bibitem[Dehnen \& Binney(1998)]{deh98} 
Dehnen, W., \& Binney, J.~J.\ 1998, \mnras, 298, 387 
\bibitem[Englmaier \& Gerhard(1999)]{eng99} 
Englmaier, P., \& Gerhard, O.\ 1999, \mnras, 304, 512 
\bibitem[Fomalont et al.(2003)]{fom03} Fomalont, E.~B., 
Petrov, L., MacMillan, D.~S., Gordon, D., \& Ma, C.\ 2003, \aj, 126, 2562 
\bibitem[Hachisuka et al.(2006)]{hac06} Hachisuka, K., et 
al.\ 2006, \apj, 645, 337 
\bibitem[Haschick \& Ho(1990)]{has90} 
Haschick, A.~D., \& Ho, P.~T.~P.\ 1990, \apj, 352, 630 
\bibitem[Honma et al.(2007)]{hon07} Honma, M., et al.\ 2007, 
\pasj, 59, 889 
\bibitem[Honma et al.(2008)]{hon08a} 
Honma, M., et al.\ 2008, \pasj, 60, 935
\bibitem[Honma et al.(2008)]{hon08b} 
Honma, M., Tamura, Y., 
\& Reid, M.~J.\ 2008, \pasj, 60, 951 
\bibitem[Iguchi et al.(2005)]{igu05} Iguchi, S., Kurayama, 
T., Kawaguchi, N., \& Kawakami, K.\ 2005, \pasj, 57, 259 
\bibitem[McMillan \& Binney(2010)]{mcm10} 
McMillan, P.~J., \& Binney, J.~J.\ 2010, \mnras, 402, 934 
\bibitem[Nagayama et al.(2008)]{nag08} Nagayama, T., 
Nakagawa, A., Imai, H., Omodaka, T., \& Sofue, Y.\ 2008, \pasj, 60, 183 
\bibitem[Pankonin et al.(2001)]{pan01} Pankonin, V., 
Churchwell, E., Watson, C., \& Bieging, J.~H.\ 2001, \apj, 558, 194 
\bibitem[Reid \& Brunthaler(2004)]{rei04} 
Reid, M.~J., \& Brunthaler, A.\ 2004, \apj, 616, 872 
\bibitem[Reid et al.(2009)]{rei09a} Reid, M.~J., et al.\ 2009, 
\apj, 700, 137 
\bibitem[Russeil(2003)]{rus03} Russeil, D.\ 2003, \aap, 397, 133 
\bibitem[Rygl et al.(2010)]{ryg09} 
Rygl, K.~L.~J., Brunthaler, A., Reid, M.~J., 
Menten, K.~M., van Langevelde, H.~J., \& Xu, Y.\ 2010, \aap, 511, A2 
\bibitem[Sato et al.(2010)]{sat10} Sato, M., Reid, M.~J., 
Brunthaler, A., \& Menten, K.~M.\ 2010, \apj, 720, 1055 
\bibitem[Sch{\"o}nrich, Binney, \& Dehnen (2010)]{sch10} 
Sch{\"o}nrich, R., Binney, J., \& Dehnen, W.\ 2010, \mnras, 403, 1829 
\bibitem[Valdettaro et al.(2001)]{val01} 
Valdettaro, R., et al.\ 2001, \aap, 368, 845 
\bibitem[Xu et al.(2006)]{xu06} Xu, Y., Reid, M.~J., Zheng, 
X.~W., \& Menten, K.~M.\ 2006, Science, 311, 54  
\bibitem[Zheng et al.(1985)]{zhe85} 
Zheng, X.~W., Ho, 
P.~T.~P., Reid, M.~J., \& Schneps, M.~H.\ 1985, \apj, 293, 522  
\end{thebibliography}
\end{document}